\documentclass[aps,prl,twocolumn,superscriptaddress,longbibliography]{revtex4-2}
\usepackage{amsmath,amssymb,amsthm,graphicx,bm,mathtools,mathrsfs}
\usepackage[hidelinks]{hyperref}
\newtheorem{theorem}{Theorem}
\newtheorem{lemma}{Lemma}
\newtheorem{smtheorem}{Theorem}
\newtheorem{smlemma}[smtheorem]{Lemma}
\newtheorem{smproposition}[smtheorem]{Proposition}
\newtheorem{smcorollary}[smtheorem]{Corollary}
\theoremstyle{remark}\newtheorem{smremark}[smtheorem]{Remark}
\theoremstyle{plain}

\newcommand{\Ltwo}{\preceq_{L^2}}
\renewcommand{\Pr}{\mathbb P}
\newcommand{\E}{\mathbb E}
\newcommand{\Tr}{{\rm Tr}\,}
\newcommand{\OG}{\mathcal O_{\rm G}}
\newcommand{\Pall}{{\rm PPT}_{\rm all}}
\newcommand{\OP}{\mathcal O_{\Pall}}
\newcommand{\SMref}[1]{[SM, sec.~\hyperref[sec:sm-#1]{#1}]}
\makeatletter
\newcommand{\equationalias}[1]{\begingroup\protected@edef\@currentlabel{\p@equation\theequation}\label{#1}\endgroup}
\makeatother

\begin{document}
\title{Reading long-range information from a separable quantum record requires entangled measurements}
\author{Qingtian Miao}
\email{qm8@tamu.edu}
\affiliation{Department of Physics and Astronomy, Texas A\&M University, Texas 77843, USA}

\begin{abstract}
Qubit probes record the relative orientation of neighboring, initially uncorrelated Ising spins through local interactions.
Each probe encodes one bond label in two states with overlap $1-p$.
The resulting probe record is fully separable.
A receiver gets one copy and guesses whether two distant spins point in the same or opposite directions.
On the square lattice, the best readout with separable measurement effects approaches success probability $1/2$
as the spin separation grows, for $0<p\le p_0=1-1/\sqrt2$.
This includes all local operations with classical communication.
The same conclusion holds for every measurement whose effects remain positive under partial transposition across every cut of the probes.
For $p_*=0.29181\le p\le1$, a joint readout of the same record maintains a success probability above $1/2$
by a positive amount independent of distance.
The overlap $[p_*,p_0]$ gives a strict separation between the readouts.
A record that is created locally and contains no entanglement can therefore hold long-range information
that only measurements with entangled effects can read.
\end{abstract}
\maketitle

\paragraph*{Introduction.}
Local interactions let a system leave records in its surroundings.
A probe qubit coupled to two neighboring spins can record their relative orientation.
Many probes can together encode information about spins far apart.
Whether this collective information can be retrieved depends on how the records are read.
A fully separable record is a mixture of product states of the probes and contains no entanglement.
It is then natural to try measuring each probe separately and combining the results classically.
Each measurement can adapt to the outcomes of earlier ones.

\begin{figure*}[t]
\includegraphics[width=\textwidth]{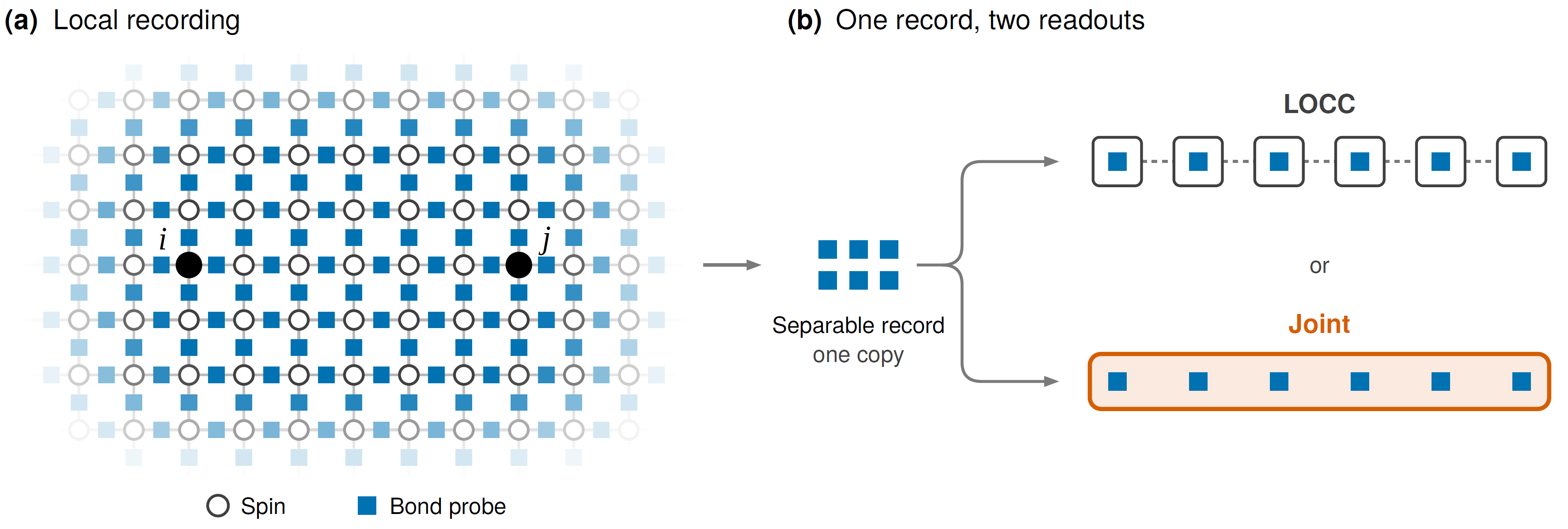}
\caption{Local recording and alternative readouts of one separable record.
(a)~Each bond probe (blue square) records the relative orientation of two neighboring spins (circles).
Averaging over the spins gives the fully separable record $\rho_A$.
The target is $T=z_iz_j$.
(b)~The same single record is read either by LOCC on separate probes or by a joint measurement.
The branches show alternative measurements on one copy.
The restricted bound also covers SEP and fully PPT measurements.}
\label{fig:story}
\end{figure*}

Even orthogonal product states can require a joint measurement for perfect discrimination~\cite{Bennett1999}.
In that example, the optimal effects are product operators.
Quantum data hiding shows that local readouts can miss an encoded bit~\cite{DiVincenzo2002}.
A bit is encoded in a pair of states that a joint measurement can distinguish well.
Local operations and classical communication (LOCC) distinguish them only poorly.
The two states can themselves be separable~\cite{Eggeling2002}.
Ha and Kim construct separable data-hiding schemes using the parity of repeated state labels~\cite{HaKim2025}.
Their constructions include two-qubit separable building blocks~\cite{HaKimTwoQubit}.
Here we study a spatially extended record generated by local spin--probe couplings.
Each physical bond of an uncorrelated Ising system leaves a record in one probe qubit with two possible pure states.
Their overlap is $1-p$, so $p$ controls how clearly the probe records its bond label.
A receiver holds one copy of all the probes and guesses whether spins $i$ and $j$ point in the same or opposite directions
(Fig.~\ref{fig:story}).
We increase the separation of $i$ and $j$ at fixed record strength $p$.

\begin{figure}[tb]
\includegraphics[width=\columnwidth]{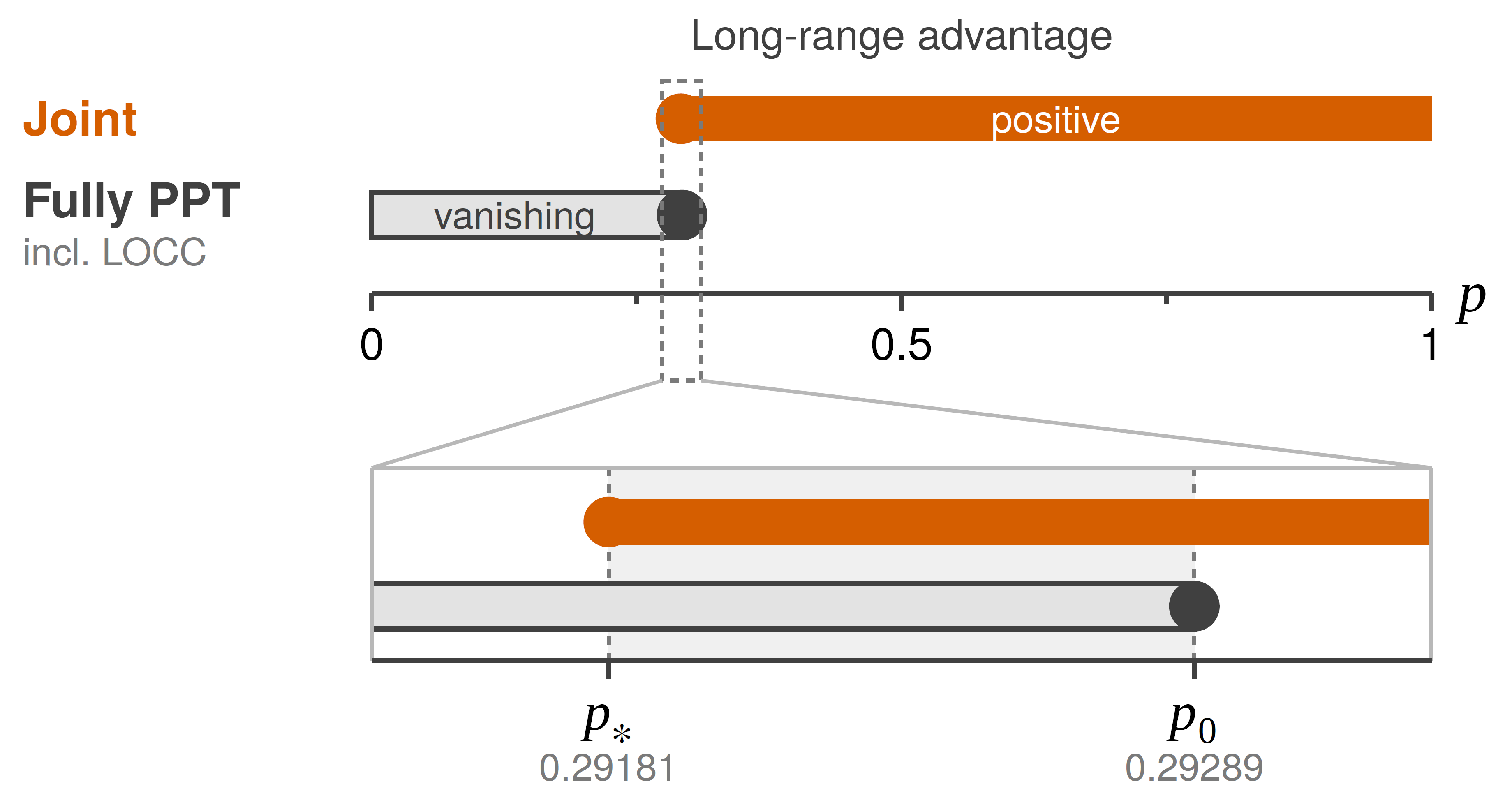}
\caption{Proved ranges of long-range prediction advantage (Theorem~\ref{thm:main}).
Joint readout has positive long-range prediction advantage on $[p_*,1]$.
Fully PPT readout, which includes SEP and LOCC, has vanishing long-range advantage on $[0,p_0]$.
Here $p=0$ is the no-record limit and $p=1$ gives orthogonal probe states.
The lower panel magnifies the certified separation window $[p_*,p_0]$,
where $p_*=0.29181$ and $p_0=1-1/\sqrt2$.
The dashed indicator box in the upper panel is widened for visibility and is not to scale.
Blank portions of each row carry no claim.}
\label{fig:result}
\end{figure}

On the square lattice we prove a strict long-range separation (Fig.~\ref{fig:result}).
For $0<p\le p_0=1-1/\sqrt2$, the best readout with separable measurement effects approaches success probability $1/2$
as the separation of $i$ and $j$ grows.
This includes every LOCC protocol, however adaptive.
For $p_*=0.29181\le p\le1$, a joint measurement of the same record keeps a positive advantage.
At all sufficiently large separations, its success probability is at least $1/2+\epsilon(p)$,
where $\epsilon(p)>0$ is independent of distance.
The two proved ranges overlap on $[p_*,p_0]$.
Within this window, any readout that keeps its advantage bounded away from zero must contain an entangled effect.
The endpoint $p_*$ is set by the present analytic lower bound.
A statistical-mechanics mapping combined with published numerics suggests a broader separation,
with the joint-readout transition near $p_G\simeq0.21844$~\cite{Lessa2025,WanDaiZhu2026}.

The restricted result holds for a larger class than separable measurements.
We allow every measurement whose effects remain positive under partial transposition (PPT) across every cut of the probes.
This class is also used to bound distinguishability in data hiding~\cite{Matthews2009,MeleLami2025,Akresh2026}.
We bound its squared prediction advantage by a random connection probability on the graph.
Edges are independently retained, and the connection event is a path joining $i$ and $j$.
This extends information-percolation bounds for classical observations of spin pairs to quantum probe records~\cite{AbbeBoix2020,PolyanskiyWu2018}.
For the joint readout, we map its prediction advantage to a classical inference problem.
This gives a positive lower bound where the restricted advantage vanishes.
Together, the two bounds establish the separation for a record generated locally with one qubit per probe.

\paragraph*{Record and task.}
Consider a finite graph $G=(V,E)$ with spin sites $V$ and bonds $E$.
A spin-$1/2$ sits at each site $v\in V$.
The number of spins is $|V|$.
The spins start in the product state $|+\rangle^{\otimes V}$, where $|+\rangle=(|0\rangle+|1\rangle)/\sqrt2$.
Let $Z_v$ be the Pauli $Z$ operator of spin $v$ and $z_v=\pm1$ its $Z$-basis label.
Every $Z$-basis configuration $z=(z_v)_{v\in V}$ has probability $2^{-|V|}$.

Each edge $e=uv$ carries one probe qubit, prepared in $\sqrt{1-p/2}\,|0\rangle+\sqrt{p/2}\,|1\rangle$.
The coupling to spins $u$ and $v$ multiplies its $|1\rangle$ amplitude by the bond label $b_e=z_uz_v$.
It can be implemented by two controlled-$Z$ gates, one from each spin to its bond probe.
For equal spin labels the probe stays unchanged.
For opposite labels the probe undergoes its Pauli phase flip $Z_e$.
The two output states are
\begin{equation}
|a_\pm\rangle=\sqrt{1-p/2}\,|0\rangle\pm\sqrt{p/2}\,|1\rangle,\qquad \langle a_+|a_-\rangle=c=1-p .
\label{eq:probe}
\end{equation}
The parameter $p\in(0,1]$ sets the record strength.
Increasing $p$ reduces the overlap $c$ and makes the two bond labels easier to distinguish.
As $p\to0$, both probe states approach $|0\rangle$ and carry no bond information.
At $p=1$, they are orthogonal and the bond label can be read without error.

For a fixed spin configuration $z$, the probes form the product state $|A_z\rangle=\bigotimes_{e\in E}|a_{b_e}\rangle$.
Tracing out the spins leaves the probe record
\begin{equation}
\rho_A=2^{-|V|}\sum_z|A_z\rangle\langle A_z| ,
\label{eq:record}
\end{equation}
a mixture of product states and hence fully separable.

The receiver's entire input is one copy of this probe record.
The task is to predict $T=z_iz_j$ for two prescribed, distinct spins $i$ and $j$.
Here $T=+1$ means the spins point in the same direction, and $T=-1$ means they point in opposite directions.
The two values are equally likely.
The records conditioned on these two values are
\begin{equation}
\rho_\pm=2^{1-|V|}\sum_{z:\,z_iz_j=\pm1}|A_z\rangle\langle A_z|.
\label{eq:conditional-record}
\end{equation}
Each sum contains $2^{|V|-1}$ spin configurations.
Both conditional records are therefore normalized mixtures of product states and are fully separable.
Predicting $T$ is the problem of distinguishing these two states with equal prior probabilities.
Write $B=\tfrac12(\rho_+-\rho_-)$ for half their difference.

A measurement has outcomes $y$ and positive effects $F_y$ that sum to the identity on the full probe space.
This is a complete positive operator-valued measure (POVM).
The conditional outcome probabilities are $P(y|T=\pm1)=\Tr\rho_\pm F_y$.
Outcome $y$ occurs with probability $P(y)=\Tr\rho_AF_y$.
For an observed $y$, Bayes' rule and the equal priors give
\begin{equation}
\begin{aligned}
P(T=\pm1|y)&=\frac{P(y|T=\pm1)P(T=\pm1)}{P(y)}\\
&=\frac{\Tr\rho_\pm F_y}{2P(y)}.
\end{aligned}
\label{eq:bayes}
\end{equation}
After seeing $y$, the receiver guesses whichever value of $T$ is more likely.
For a measurement class $\mathcal M$, let $P_{\rm guess}^{\mathcal M}$ be the best achievable success probability.
The prediction advantage is
\begin{equation}
\mathcal O_{\mathcal M}(i,j)=\sup_{\{F_y\}\in\mathcal M}\sum_yP(y)\,\big|\E[T|y]\big|=2P_{\rm guess}^{\mathcal M}-1 .
\label{eq:O}
\end{equation}
Here $\E[T|y]=P(T=+1|y)-P(T=-1|y)$ measures which value of $T$ is favored by outcome $y$.
Bayes' rule therefore expresses the advantage in terms of the conditional-state difference,
\begin{equation}
\begin{aligned}
P(y)\E[T|y]&=\Tr BF_y,\\
\mathcal O_{\mathcal M}(i,j)&=\sup_{\{F_y\}\in\mathcal M}\sum_y|\Tr BF_y|.
\end{aligned}
\label{eq:score}
\end{equation}
Every outcome contributes with its probability $P(y)$.
Thus the advantage measures how well class $\mathcal M$ distinguishes $\rho_+$ from $\rho_-$.
The two states and their prior probabilities are the same for all classes.

The measurement partition has one qubit per probe.
GLOBAL (G) allows every complete POVM on these probes.
LOCC allows local operations and classical communication, with one laboratory per probe.
It includes adaptive rounds, local auxiliary systems and local quantum memory.
SEP requires each effect to be fully separable over the probes, a sum of positive product operators.
Every LOCC protocol has this property~\cite{Chitambar2014}.
The larger class $\Pall$ requires $F_y^{T_S}\succeq0$ for every outcome $y$ and every subset $S$ of probes.
Here $T_S$ transposes the matrix indices of the probes in $S$ in their $|0\rangle,|1\rangle$ bases.
We call this class fully PPT because its effects remain positive under partial transposition across every cut.
The measurement classes satisfy
\begin{equation}
{\rm LOCC}\subseteq{\rm SEP}\subseteq\Pall\subseteq{\rm GLOBAL}.
\label{eq:classes}
\end{equation}

For GLOBAL, measuring in an eigenbasis of $B$ attains $\OG=\|B\|_1$.
The trace norm is the sum of the absolute eigenvalues of $B$.

On the square lattice we use free-boundary boxes $\Lambda_L=\{-L,\dots,L\}^2$ with $i=0$, $j=r\hat x$, $L\ge r$.
Only bonds with both vertices inside the box are included.
Here $\hat x$ is a lattice unit vector and $r$ is the spin separation in lattice spacings.
Write $\mathcal O_{\mathcal M,L}$ for the advantage in this box and define
\begin{equation}
\mathcal O_{\mathcal M}(r)=\lim_{L\to\infty}\mathcal O_{\mathcal M,L}(0,r\hat x).
\label{eq:distance}
\end{equation}
Enlarging the box adds probes that the receiver may ignore, so this bounded advantage has a limit~\SMref{S1}.
We take this system-size limit at each fixed $r$ before taking the large-distance limit.

\paragraph*{Main result.}
The receiver gets one copy of the same fully separable probe record and guesses whether two distant spins point
in the same or opposite directions. Only the allowed measurement of the probes changes.
\begin{theorem}\label{thm:main}
Let $p_0=1-1/\sqrt2\approx0.2929$ and $p_*=0.29181$. Consider the square-lattice record with one qubit per probe.
Use the measurement classes defined above. First take $L\to\infty$ at fixed $r$, then take $r\to\infty$. We have
\begin{equation}
\begin{aligned}
\lim_{r\to\infty}\OP(r)&=0 &&(0<p\le p_0),\\
\liminf_{r\to\infty}\OG(r)&>0 &&(p_*\le p\le1).
\end{aligned}
\label{eq:main}
\end{equation}
The first conclusion also holds for SEP and LOCC. For the second, we prove a positive lower bound at $p=p_*$.
The joint advantage is nondecreasing in $p$, so the bound holds throughout $[p_*,1]$.
The two ranges overlap on $[p_*,p_0]$. This is the strict separation window.
\end{theorem}
The success probability is $P_{\rm guess}^{\mathcal M}=(1+\mathcal O_{\mathcal M})/2$.
In the separation window, the best $\Pall$, SEP and LOCC readouts approach success probability $1/2$.
Guessing the spins' relative orientation becomes no better than a fair coin toss.
A joint readout of the same record succeeds with probability at least $1/2+\epsilon(p)$ at all sufficiently large distances.
Here $\epsilon(p)>0$ is independent of distance.
Any measurement family that retains such an advantage must contain an entangled effect at all sufficiently large distances.
At least one effect has a negative partial transpose across some cut of the probes.
The cut may depend on distance.

For $p\in[p_*,p_0]$, Eq.~\eqref{eq:main} has the equivalent finite-box form
\begin{equation}
\begin{gathered}
\limsup_{r\to\infty}\,\sup_{L\ge r}\,\mathcal O_{\Pall,L}(0,r\hat x)=0,\\
\liminf_{r\to\infty}\,\liminf_{L\to\infty}\,\mathcal O_{{\rm G},L}(0,r\hat x)>0.
\end{gathered}
\label{eq:main-finite}
\end{equation}
The system-size limit is taken at each fixed distance before the distance limit.

\begin{figure*}[t]
\includegraphics[width=\textwidth]{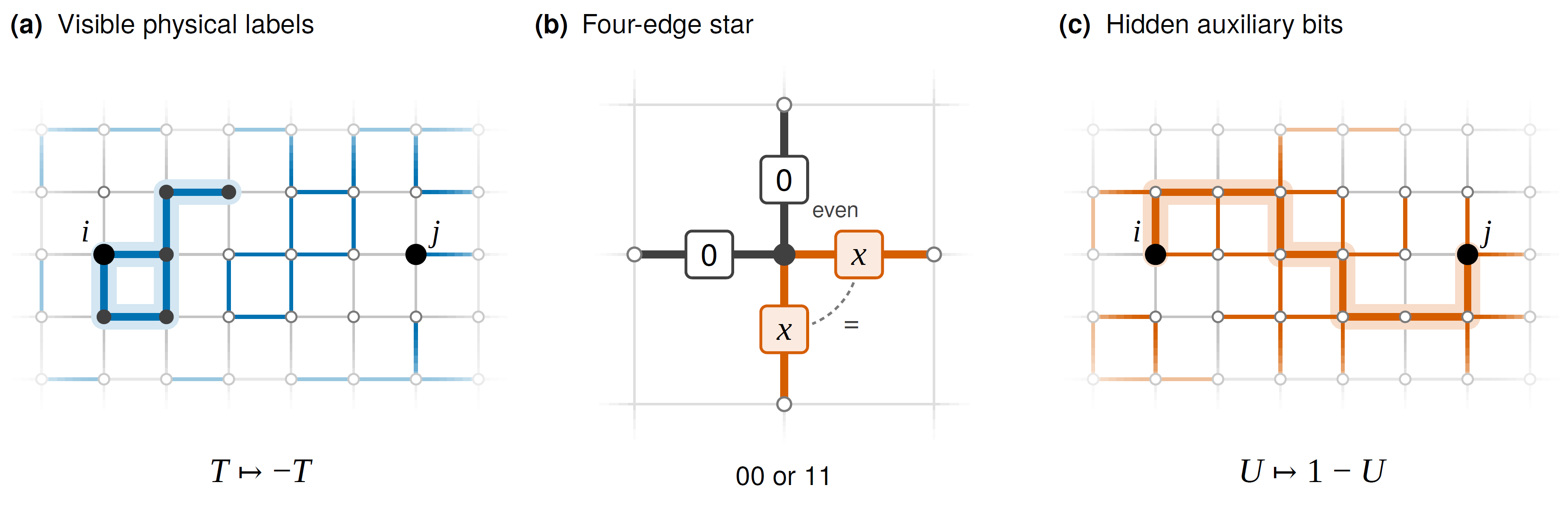}
\caption{Physical reveal clusters and auxiliary hidden paths in the comparison models (schematic).
(a)~The blue edges carry revealed physical bond labels.
Flipping every spin in the highlighted component containing $i$ preserves those labels and reverses $T$.
(b)~On a four-edge auxiliary star away from $i$ and $j$, even parity forces the two hidden bits to agree
when the two revealed bits are both zero.
The remaining uncertainty is a single bit repeated on two edges.
(c)~Orange marks hidden auxiliary edges regardless of their bit values.
Flipping the bits along the highlighted $i$--$j$ path preserves the revealed values and changes $U$ to $1-U$.
At $p=p_*$, the four-edge channel comparison gives independent interior hidden edges with probability
$1-\tau_*=\tfrac12+10^{-7}$, while the physical reveal probability is $s_*=p_*(2-p_*)<\tfrac12$.
The factor $s_*^8$ in the joint lower bound accounts for requiring up to eight edges
in exceptional stars at $i$ or $j$ to be hidden.}
\label{fig:mechanism}
\end{figure*}

Both bounds use random paths on the lattice [Fig.~\ref{fig:mechanism}].
In bond percolation, each edge is independently kept with a specified probability.
One asks whether the kept edges form a path between $i$ and $j$.
For the restricted bound, we imagine independently revealing some physical bond labels $b_e=z_uz_v$.
The advantage is at most the square root of the probability that these revealed edges join $i$ and $j$.

For the joint bound, we introduce a classical inference problem on the same lattice.
Each edge carries an auxiliary bit.
The number of bits equal to 1 meeting at a vertex is even, except possibly at $i$ and $j$.
The task is to infer whether $i$ and $j$ have odd parity from noisy observations of the edge bits.
The harder this auxiliary task is, the larger the physical joint readout advantage.
To bound the joint advantage, we compare noisy observations with observations that either reveal or hide each bit.
The probability that hidden edges join $i$ and $j$ is then a lower bound on the joint advantage.

Comparing one edge at a time gives the same percolation threshold for both bounds.
It therefore gives no separation window.
We instead use the four edges meeting at a vertex, called a star.
Away from $i$ and $j$, their auxiliary bits have even parity.
Once two bits are known, this parity constraint ties the remaining two together.
Comparing all four bits together allows a larger probability of hiding each bit.
The hidden edges can then join $i$ and $j$ at large separation.
Their connection probability stays bounded away from zero at $p=p_*$.
Monotonicity in record strength extends the joint lower bound to the separation window.
We now explain both bounds.

\paragraph*{Why restricted readouts fail.}
Consider a classical comparison record that reveals each physical bond label $b_e=z_uz_v$ independently with probability $s=p(2-p)$.
Otherwise, the output marks that label as missing.
This is an erasure, and its location is known.
Let $\mathcal R$ contain the revealed labels and the set of edges on which they were revealed.
If a revealed path joins $i$ and $j$, multiplying its labels gives $T=z_iz_j$.
If there is no such path, consider the cluster of vertices connected to $i$ by revealed edges.
Flipping every spin in this cluster leaves all revealed labels unchanged and reverses $T$.
The two spin configurations have equal prior probability.
Thus the revealed record determines $T$ when $i$ and $j$ are connected and leaves it completely unknown otherwise.
\begin{equation}
\E[T|\mathcal R]^2=\bm 1\{i\leftrightarrow j\}.
\label{eq:revealed-parity}
\end{equation}
Here $\E[T|\mathcal R]$ averages $T$ over spin configurations consistent with the revealed record.
The notation $i\leftrightarrow j$ means that a revealed path joins $i$ and $j$.
The symbol $\bm 1$ equals 1 when this happens and 0 otherwise.

The revealed edges form independent bond percolation with edge probability $s$.
Write $\Pr_s(i\leftrightarrow j)$ for the probability that they connect $i$ and $j$.
This probability bounds the squared advantage of every $\Pall$ readout.
\begin{theorem}\label{thm:ppt}
On every finite graph with one qubit per bond, the optimum over complete $\Pall$ POVMs obeys, for all $i\neq j$ and $0<p\le1$,
\begin{equation}
\OP(i,j)^2\le\Pr_s(i\leftrightarrow j),\qquad s=p(2-p).
\label{eq:ppt}
\end{equation}
\end{theorem}
The PPT condition makes the connection bound apply to the quantum probes.
Combine outcomes giving the same guess into two effects $F,I-F$ and set $H=2F-I$.
Average $H$ over partial transposes on all subsets of probes to obtain $K$.
The fully PPT condition keeps each transposed observable, and hence $K$, between $-I$ and $I$.
At every fixed spin configuration, the record is a real product state.
The averaging therefore preserves the score and removes every Pauli product containing a local $Y$.
The remaining $I,X,Z$ products have a classical representation with the same first and second moments.
Let $\mathcal A=\Tr BH=\Tr BK$ be the readout advantage.
For $0<p<1$, one can associate a real function $k(\mathcal R)$ with this readout such that
\begin{equation}
\E[Tk(\mathcal R)]=\mathcal A,\qquad \E[k(\mathcal R)^2]\le1 .
\label{eq:rep}
\end{equation}
The spectral bound on $K$ controls this second moment~\SMref{S1}.
Combining Eqs.~\eqref{eq:revealed-parity} and \eqref{eq:rep}, Cauchy--Schwarz gives
\begin{equation}
\begin{aligned}
\mathcal A^2&=\big(\E[\E[T|\mathcal R]k(\mathcal R)]\big)^2\\
&\le\E[\E[T|\mathcal R]^2]\,\E[k(\mathcal R)^2]\\
&\le\Pr_s(i\leftrightarrow j).
\end{aligned}
\label{eq:ppt-moments}
\end{equation}
This holds for every allowed readout and proves Eq.~\eqref{eq:ppt}.
Physically, the receiver loses its advantage when revealed paths between distant spins become rare.
At $p=1$, every physical bond label is perfectly readable and the same connection argument applies.

Trees give a useful reference point.
The unique path between $i$ and $j$ still carries information about their relative orientation.
Suppose the path has $d$ edges.
Measure each probe in the basis $(|0\rangle\pm|1\rangle)/\sqrt2$.
This is an $X$ measurement and reaches the joint optimum,
\begin{equation}
\OG(i,j)=\OP(i,j)=\mathcal O_{\rm LOCC}(i,j)=s^{d/2}.
\label{eq:tree-path}
\end{equation}
For $0<p<1$, the advantage is positive at every finite distance and decays as the path grows.
Thus a separation between the readouts requires loops.
For $0<p<1$, a triangle already has a joint advantage above the $\Pall$ optimum.
A protocol using one round of classical communication attains the restricted optimum~\SMref{S1}.
This is a finite-graph example of how a loop helps the joint readout.
On the square lattice, we use the connection bound to determine when the restricted advantage vanishes.

On the infinite square lattice, the vertices reachable from the origin along revealed edges form its cluster.
An infinite cluster extends to arbitrarily large distances.
Write $\theta(s)$ for the probability that the origin belongs to such a cluster.
For $s\le1/2$, no infinite revealed cluster forms, so $\theta(s)=0$.
This includes the critical value $s=1/2$~\cite{Harris1960,BollobasRiordan2006}.
To connect the origin to a spin at distance $r$, its revealed cluster must extend at least that far.
The probability of this tends to zero as $r$ grows.
This bounds the connection probability in every finite box, independently of the box size.
We can therefore take the system-size limit first and the distance limit afterward.
Since $s\le1/2$ exactly when $p\le p_0$, Eq.~\eqref{eq:ppt} gives the restricted conclusion of Eq.~\eqref{eq:main}.
For $p<p_0$, the decay with distance is exponential~\cite{AizenmanBarsky1987}.

\paragraph*{Why the joint readout succeeds.}
Let $\rho_S$ be the spin state after the interaction with the probes.
The joint advantage is its fidelity overlap with the state obtained by applying $Z_iZ_j$~\cite{Lessa2025},
\begin{equation}
\OG=\|B\|_1=F(\rho_S,Z_iZ_j\rho_SZ_iZ_j),
\label{eq:fid}
\end{equation}
where $F(\rho,\sigma)=\|\sqrt\rho\sqrt\sigma\|_1$ is the root fidelity.
The operators $Z_i$ and $Z_j$ reverse the $X$-basis spin labels at $i$ and $j$.

We map the joint readout to a classical inference problem on the same lattice~\SMref{S3}.
Assign an auxiliary bit $x_e\in\{0,1\}$ to each edge and mark the edges with $x_e=1$.
Choose a bit $U$ to be 0 or 1 with equal probability.
For $U=0$, an even number of marked edges meet at every vertex.
The marked edges can then be decomposed into closed loops.
For $U=1$, the count is odd at $i$ and $j$ and even at every other vertex.
The marked edges then consist of a path from $i$ to $j$ together with closed loops.
For each value of $U$, choose uniformly among all configurations obeying its counting rule.
The two choices of $U$ select the vertex-parity sectors related by $Z_iZ_j$ in the spin fidelity.
Related inference problems with noisy edge bits and vertex-parity constraints occur in surface-code decoding~\cite{Dennis2002}.

The noise in this classical problem comes from the spin dephasing.
Tracing out the probe on bond $uv$ gives the following channel on a spin state $\rho$,
\begin{equation}
\mathcal D_{uv}(\rho)=(1-q)\rho+qZ_uZ_v\rho Z_uZ_v,\qquad q=p/2.
\label{eq:dephasing}
\end{equation}
Thus each bond independently applies $Z_uZ_v$ with probability $q$.
In the $X$ basis, each such operation flips the spin labels at $u$ and $v$.
In the classical mapping, these random bond events become independent flips of the edge bits $x_e$.
The observed bits $y=\{y_e\}$ are therefore $x_e$ flipped with probability $q$ on each edge.
The task is to infer $U$ from $y$.
As $p$ grows, the physical probe states become easier to distinguish and the auxiliary bits suffer more flips.

The overlap of the output distributions for $U=0$ and $U=1$ equals the spin-state fidelity in Eq.~\eqref{eq:fid}.
After observing $y$, write $m(y)=P(U=0|y)-P(U=1|y)$.
Then
\begin{equation}
\OG=\E_y\sqrt{1-m(y)^2}\ \ge\ 1-\E_y[m(y)^2] .
\label{eq:aux}
\end{equation}
If $y$ leaves both values of $U$ equally likely, $m(y)=0$ and the overlap factor is 1.
If $y$ determines $U$, $|m(y)|=1$ and the factor is 0.
Thus hiding $U$ in the auxiliary problem certifies an advantage in the physical joint readout.

Compare each noisy observation with one that reveals its bit exactly with probability $c^2$, where $c=1-p$.
Otherwise the bit is marked as missing.
This comparison can only increase $\E m^2$ and hence gives a lower bound in Eq.~\eqref{eq:aux}~\cite{Nair2009,MakurPolyanskiy2018}.
Figure~\ref{fig:mechanism}(c) shows a hidden path between $i$ and $j$.
Orange marks hidden positions, regardless of whether their bits are 0 or 1.
Flipping every bit along this path leaves all revealed values unchanged.
At each intermediate vertex, two path bits flip, preserving the even-parity rule.
At $i$ and $j$, only one bit flips, so $U$ changes to $1-U$.
The two configurations have equal prior probability, leaving both values of $U$ equally likely.
Without a hidden path, the revealed bits and the vertex constraints determine $U$.
Thus
\begin{equation}
\OG\ge\Pr_s(i\leftrightarrow j),\qquad s=1-c^2.
\label{eq:global-edgewise}
\end{equation}
Here the paths use hidden auxiliary edges.
The lower bound guarantees long-range advantage for $s>1/2$.
The restricted bound rules it out for $s\le1/2$.
These ranges meet at $p_0$ and give no separation window.

At $p=p_*$, we instead compare the four edges meeting at one vertex, called a \emph{star}.
At vertices other than $i$ and $j$, their auxiliary bits contain an even number of 1s.
Figure~\ref{fig:mechanism}(b) shows two revealed bits equal to 0.
The hidden pair must then be either 00 or 11.
Knowing two bits fixes whether the other two are equal or opposite.
For opposite bits, relabeling one as its complement gives two equal copies.
The remaining pair thus carries one bit twice, up to this known relabeling.
Two noisy copies can agree and still give the wrong value.
An exact reveal fixes that value as soon as either copy is visible.
This repeated-bit gain lowers the required reveal probability to
\begin{equation}
\tau_*=\tfrac12-10^{-7}.
\label{eq:main-star-rate}
\end{equation}
The star comparison at this rate can only increase $\E m^2$~\SMref{S3}.

Color the vertices black and white and group the edges around each black vertex.
The comparison holds for every allowed distribution of the four bits.
It can therefore be applied star by star, even when observations elsewhere change those probabilities.
The observation noise on different stars is independent.

The resulting hidden edges have probability $1-\tau_*>1/2$ in the interior.
They form an infinite cluster~\cite{Kesten1980}.
Uniqueness of the infinite cluster and the Harris--FKG inequality keep the connection probability between $i$ and $j$ bounded away from zero~\cite{AKN1987,BurtonKeane1989,Harris1960,Fortuin1971}.
A finite correction near $i$ and $j$ gives a distance-independent factor $s_*^8$, where $s_*=p_*(2-p_*)$.
Taking the system-size limit at fixed distance and then the distance limit gives, at $p=p_*$,
\begin{equation}
\liminf_{r\to\infty}\OG(r)\ \ge\ s_*^8\,\theta(1-\tau_*)^2>0.
\label{eq:global}
\end{equation}
Here $\theta(1-\tau_*)$ is the probability that a vertex belongs to the infinite cluster of hidden edges.

A local operation on each probe can turn a stronger record into a weaker one.
The receiver can apply these operations before reading the probes.
Hence $\OG$ is nondecreasing in $p$, and the lower bound holds throughout $[p_*,1]$.
Its overlap with the restricted range gives the separation window $[p_*,p_0]$.
This completes Theorem~\ref{thm:main}.

\paragraph*{Discussion.}
Local interactions can create a fully separable record whose long-range information requires entangled measurement effects.
In the separation window, a receiver restricted to separable effects loses the ability to predict whether distant spins are aligned.
A joint readout of the same single record retains a success probability above random guessing by a distance-independent amount.
The entanglement needed to read this information is a property of the measurement effects.

Loop constraints organize correlations among the probes.
On any set of bonds containing no closed loop, the bond labels are independent and uniform.
After averaging over the spins, the reduced record on those bonds is therefore a product state~\SMref{S3}.
This includes any pair of probes on the square lattice.
The prediction task depends on the records conditioned on $T$.
The product of bond labels along a path from $i$ to $j$ equals $T$, so fixing $T$ constrains their values together.
Such paths carry predictive information even on a tree.
There, measuring each probe along the path already attains the joint optimum in Eq.~\eqref{eq:tree-path}.
Loops are needed for a separation between the readouts.

The width of the certified interval reflects the reach of the four-edge comparison.
The statistical-mechanics description of the same spin state points to a wider separation region.
Lessa et al. mapped its fidelity correlator to a classical square-lattice Ising model with random coupling signs~\cite{Lessa2025}.
Negative couplings represent the independent dephasing events and occur with probability $q=p/2$.
The effective dimensionless coupling $\mathcal K$ obeys $e^{-2\mathcal K}=q/(1-q)$.
This matching of temperature and disorder defines the Nishimori line~\cite{Nishimori1981}.
The operation $Z_iZ_j$ inserts a defect line joining $i$ and $j$ by reversing coupling signs.
Let $\mathcal Z$ and $\mathcal Z_{ij}$ be the Ising partition functions without and with this defect.
Its dimensionless free-energy cost gives
\begin{equation}
\begin{aligned}
\Delta\mathcal F_{ij}&=-\ln(\mathcal Z_{ij}/\mathcal Z),\\
\OG(i,j)&=\E_q\big[e^{-\Delta\mathcal F_{ij}/2}\big].
\end{aligned}
\label{eq:domain-wall}
\end{equation}
Here $\E_q$ averages over the independent random coupling signs~\SMref{S3}.
In the disordered Ising phase, the mean defect cost is expected to stay bounded as the separation grows.
This would preserve an overlap bounded away from zero between the two spin states.
By Eq.~\eqref{eq:fid}, this overlap is the receiver's joint prediction advantage.

Wan, Dai and Zhu located the Nishimori transition using fermionic transfer matrices and finite-size scaling~\cite{WanDaiZhu2026}.
Their largest lattices were $512\times512$, with more than $10^7$ disorder realizations.
Their bond-disorder probability is our dephasing rate $q$.
In our notation, their estimate and its conversion are
\begin{equation}
q_N=0.1092212(4),\qquad p_G=2q_N=0.2184424(8).
\label{eq:nishimori-point}
\end{equation}
The factor of two follows from Eq.~\eqref{eq:dephasing}.
Together with the proved restricted bound, this estimate suggests separation throughout $p_G<p\le p_0$.
The numerically predicted interval has width about $0.074$, compared with about $0.0011$ for the certified interval.
The analytic result certifies a portion of this broader region suggested by the numerics.

The spin fidelity is also an order parameter for strong-to-weak spontaneous symmetry breaking (SWSSB)~\cite{Lessa2025}.
Here it has a direct operational meaning as the best prediction advantage available from one copy of the probe record.
General SWSSB detection asks how efficiently this order can be identified in an unknown state~\cite{Feng2025}.
Our result concerns the measurement effects needed to read a known record.
Local ancilla readout also appears in measurement-based preparations of Nishimori-type states~\cite{Zhu2023,Chen2025}.
When those ancillas realize our record and probe partition, Theorem~\ref{thm:ppt} bounds their local readout.

A natural next step is to construct joint readouts that retain the long-range advantage.
Such protocols could reveal how the advantage depends on the number of coherently measured probes and the circuit depth.

\clearpage
\onecolumngrid
\begingroup
\setcounter{equation}{0}
\setcounter{table}{0}
\setcounter{section}{0}
\setcounter{smtheorem}{0}
\renewcommand{\theequation}{S\arabic{equation}}
\renewcommand{\theHequation}{S\arabic{equation}}
\renewcommand{\theHsection}{S\arabic{section}}
\renewcommand{\thetable}{S\Roman{table}}
\let\theorem\smtheorem\let\endtheorem\endsmtheorem
\let\lemma\smlemma\let\endlemma\endsmlemma
\let\proposition\smproposition\let\endproposition\endsmproposition
\let\corollary\smcorollary\let\endcorollary\endsmcorollary
\let\remark\smremark\let\endremark\endsmremark

\begin{center}
{\large\bfseries Supplemental Material}\\[4pt]
{\bfseries Reading long-range information from a separable quantum record requires entangled measurements}
\end{center}

The author used GPT-6 Astra (OpenAI) to explore proof strategies.
The author directed this work and verified the resulting arguments.
The author takes responsibility for the final results and their presentation.

We use the probe record and spin-parity task defined in the main text.
Section S1 compares every fully PPT readout with a classical record of randomly revealed bond labels.
Section S2 extends this bound to the uncertainty remaining after each measurement outcome.
Section S3 obtains the joint lower bound from hidden paths in the auxiliary problem.
We keep the main-text notation, including $c=1-p$, $s=1-c^2$ and $q=p/2$.
The square-lattice boxes and order of limits are those of Eq.~\eqref{eq:distance}.

\section{S1. The all-cuts PPT bound (Theorem~\ref{thm:ppt})}\label{sec:sm-S1}

\subsection{The record and its prediction score}
The receiver distinguishes the two conditional records $\rho_+$ and $\rho_-$ with equal prior probabilities.
For a fixed spin configuration, write $\rho_z=|A_z\rangle\langle A_z|$.
The phase-flip preparation used in the main text is also equivalent to a controlled probe rotation.
Preparing $|0\rangle$ and applying $e^{-i\vartheta Z_uZ_vY_e/2}$ gives $|a_{b_e}\rangle$ when $\sin^2(\vartheta/2)=p/2$ and $0<\vartheta\le\pi/2$.
Here $Y_e$ is Pauli $Y$ on probe $e$.
Both preparations give the same state for every spin configuration.
The operator that weights each record by its target parity is
\begin{equation}
B=\tfrac12(\rho_+-\rho_-)=2^{-|V|}\sum_zT(z)\rho_z.
\label{eq:B-definition}
\end{equation}
Equation~\eqref{eq:score} expresses the prediction advantage as a distinguishability norm~\cite{Matthews2009},
\begin{equation}
\|B\|_{\mathcal M}=\sup_{\{F_y\}\in\mathcal M}\sum_y|\Tr BF_y|=\mathcal O_{\mathcal M}(i,j).
\label{eq:distinguishability}
\end{equation}
For unrestricted measurements, the Holevo--Helstrom theorem gives $\OG=\|B\|_1$~\cite[Theorem~3.4]{Watrous2018}.
We now bound the same score for fully PPT measurements.
All effects act on the full probe space $\mathcal H=\bigotimes_{e\in E}\mathbb C^2$, with one qubit per probe.
The measurement classes are defined in the main text.
The inclusion of LOCC in SEP follows from the standard characterization of LOCC measurements~\cite{Chitambar2014}.

Group all outcomes favoring $T=+1$ into one effect $F$.
The other effect is $I-F$, which gives the guess $T=-1$.
Adding effects preserves both separability and positivity under every partial transpose.
The binary observable $H=2F-I$ therefore gives the same prediction advantage as the original readout,
\begin{equation}
\mathcal O_{\mathcal M}(i,j)
=\sup_{\{F,I-F\}\in\mathcal M}\Tr BH,\qquad H=2F-I.
\label{eq:binary-readout}
\end{equation}
\equationalias{lem:two}
Here $\Tr B=0$, so the score is $2\Tr BF$.
Exchanging the two guesses changes its sign.

\subsection{Keeping the score while removing local Pauli \texorpdfstring{$Y$}{Y} terms}
To compare the quantum score with revealed bond information, we first simplify its observable.
Let $M=|E|$ be the number of probes.
Average $H$ over transposition of every possible subset $S$ of probes,
\begin{equation}
K=2^{-M}\sum_{S\subseteq E}H^{T_S}.
\label{eq:transpose-average}
\end{equation}
For a fully PPT binary measurement, both $F^{T_S}$ and $(I-F)^{T_S}$ are positive.
Thus every $H^{T_S}$ has eigenvalues between $-1$ and $1$.
Their average has the same property.
Each $\rho_z$ is a product of real single-probe states, so transposing any subset leaves it unchanged.
This gives $B^{T_S}=B$ and $\Tr BH^{T_S}=\Tr BH$.
Consequently,
\begin{equation}
-I\preceq K\preceq I,\qquad K^2\preceq I,\qquad \Tr BK=\Tr BH.
\label{eq:average-properties}
\end{equation}
\equationalias{lem:avg}
The matrix order $A\preceq C$ means that $C-A$ is positive semidefinite.
The last identity preserves the prediction score.
The bound on $K^2$ will control the fluctuations in the classical comparison.

A local transpose fixes $I,X,Z$ and changes $Y$ to $-Y$.
In Eq.~\eqref{eq:transpose-average}, terms containing $Y$ on any probe therefore cancel in pairs.
The remaining operator $K$ is a real linear combination of products of $I,X,Z$ factors.
This averaging is a step in the proof.
The receiver's actual measurement is still the original fully PPT readout.

\subsection{Matching the score and its second moment with revealed labels}
Imagine revealing each physical bond label independently with probability $s$.
An unrevealed label is replaced by a missing-label mark, so its location is known.
Call the complete revealed record $\mathcal R$.
We will represent the score of $K$ by a real function $k(\mathcal R)$.
The construction matches both its correlation with $T$ and its second moment.

Take $0<p<1$ for this construction.
For one probe carrying label $t=\pm1$,
\begin{equation}
\omega_t=|a_t\rangle\langle a_t|=\tfrac12(I+\sqrt s\,tX+cZ).
\label{eq:single-probe-bloch}
\end{equation}
Let $R_e=1$ mean that the label on edge $e$ is revealed, and $R_e=0$ mean that it is missing.
The indicators $R_e$ are independent of the spins and of one another, with $\Pr(R_e=1)=s$ and $\Pr(R_e=0)=c^2$.
Assign a real function to each of the three Pauli factors,
\begin{center}
\begin{tabular}{c|cc}
 & revealed label $t$ & missing label\\ \hline
$f_I$ & $1$ & $1$\\
$f_X$ & $t/\sqrt s$ & $0$\\
$f_Z$ & $0$ & $1/c$
\end{tabular}
\end{center}
These functions reproduce the single-probe moments,
\begin{equation}
\E_R[f_af_b\,|\,t]=\Tr\omega_t ab,\qquad a,b\in\{I,X,Z\}.
\label{eq:moment}
\end{equation}
\equationalias{lem:mom1}
Here $\E_R$ averages over whether the label is revealed.
For example, $\E_R f_X=\sqrt s\,t$ and $\E_R f_Z=c$, while $\E_R f_X^2=\E_R f_Z^2=1$.
The mixed product $f_Xf_Z$ is zero, matching the zero expectation of $XZ$ and $ZX$ in the real state $\omega_t$.
These checks also give the cases involving $I$.

Write the Pauli expansion of $K$ as
$K=\sum_{\boldsymbol\alpha}k_{\boldsymbol\alpha}\bigotimes_{e\in E}\alpha_e$,
where $\boldsymbol\alpha$ specifies one factor $\alpha_e\in\{I,X,Z\}$ on each probe.
Replacing every factor by its function from the table defines
\begin{equation}
k(\mathcal R)=\sum_{\boldsymbol\alpha}k_{\boldsymbol\alpha}\prod_{e\in E}f_{\alpha_e}(R_e,b_e).
\label{eq:revealed-score-function}
\end{equation}
At fixed $z$, the probes form a product state and all reveals are independent.
The single-probe identity therefore multiplies across the edges for every pair of Pauli products.
Expanding $K$ and $K^2$ gives
\begin{equation}
\begin{aligned}
\E_R[k(\mathcal R)|z]&=\Tr\rho_zK,\\
\E_R[k(\mathcal R)^2|z]&=\Tr\rho_zK^2.
\end{aligned}
\label{eq:conditional-revealed-moments}
\end{equation}
\equationalias{lem:mom}
Averaging over the uniformly distributed spin configurations then yields
\begin{equation}
\E[Tk(\mathcal R)]=\Tr BK,\qquad
\E[k(\mathcal R)^2]=\Tr\rho_AK^2\le1.
\label{eq:revealed-moment-bound}
\end{equation}
The function $k$ can take values outside $[-1,1]$.
Its role is to reproduce the score, with the second moment providing the bound needed below.
The revealed record is a mathematical comparison, rather than an additional input to the receiver.

\subsection{Revealed paths and the proof of Theorem~\ref{thm:ppt}}
A revealed path from $i$ to $j$ determines $T$ by multiplying the bond labels along it.
When there is no such path, consider all sites reachable from $i$ along revealed edges.
These sites form the revealed cluster of $i$, illustrated in Fig.~\ref{fig:mechanism}(a).
Flip every spin in this cluster.
Every revealed bond has either both sites flipped or neither site flipped, so all revealed labels stay the same.
Spin $j$ lies outside the cluster, so this operation reverses $T$.
The original and flipped configurations have equal prior probability.
Thus $T=+1$ and $T=-1$ remain equally likely after seeing $\mathcal R$.

Write $i\leftrightarrow j$ for connection by revealed edges.
Each edge is revealed independently with probability $s$, so these edges form independent bond percolation.
The two cases above give
\begin{equation}
\begin{aligned}
\E[T|\mathcal R]^2&=\bm 1\{i\leftrightarrow j\},\\
\E\big[\E[T|\mathcal R]^2\big]&=\Pr_s(i\leftrightarrow j).
\end{aligned}
\label{eq:revealed-connectivity}
\end{equation}
\equationalias{lem:perc}
The indicator in the first line is 1 when the sites are connected and 0 otherwise.

For any fully PPT readout, Eqs.~\eqref{eq:binary-readout} and \eqref{eq:average-properties} express its advantage as $\Tr BK$.
Using the moment comparison in Eq.~\eqref{eq:revealed-moment-bound}, Cauchy--Schwarz gives
\begin{equation}
\begin{aligned}
(\Tr BK)^2
&=\big(\E[\E[T|\mathcal R]k(\mathcal R)]\big)^2\\
&\le\E\big[\E[T|\mathcal R]^2\big]\,\E[k(\mathcal R)^2]\\
&=\Pr_s(i\leftrightarrow j)\,\Tr\rho_AK^2\\
&\le\Pr_s(i\leftrightarrow j).
\end{aligned}
\label{eq:ppt-proof}
\end{equation}
Taking the best readout proves Theorem~\ref{thm:ppt}.
Physically, the restricted score can remain large only while the revealed comparison record connects the two spins with appreciable probability.
At $p=1$, each probe reveals its bond label without error.
A connected graph then determines $T$ exactly.
If $i$ and $j$ lie in different graph components, flipping all spins in one component preserves the entire record and reverses $T$.
The same bound therefore holds at $p=1$.

\subsection{Loss of long-range prediction on the square lattice}
First consider increasing the box size at a fixed spin separation $r$.
The joint distribution of $T$ and the probes in the smaller box stays unchanged.
The receiver may ignore the added probes by replacing each effect $F_y$ with $F_y\otimes I$.
This preserves GLOBAL, LOCC, SEP and fully PPT measurements.
The optimal advantage is therefore nondecreasing with box size and bounded by 1.
Consequently,
\begin{equation}
\mathcal O_{\mathcal M}(r)
=\lim_{L\to\infty}\mathcal O_{\mathcal M,L}(0,r\hat x)
=\sup_{L\ge r}\mathcal O_{\mathcal M,L}(0,r\hat x).
\label{eq:box-limit}
\end{equation}

A revealed cluster must extend at least a distance $r$ to connect $0$ to $r\hat x$.
Let $\partial\Lambda_r$ be the sites on the four outer sides of $\Lambda_r$, where $\max(|x|,|y|)=r$.
The infinite square lattice is denoted by $\mathbb Z^2$.
Adding the bonds outside a finite box can only help connection, so for every $L\ge r$,
\begin{equation}
\Pr_{s,\Lambda_L}(0\leftrightarrow r\hat x)
\le\Pr_{s,\mathbb Z^2}(0\leftrightarrow\partial\Lambda_r).
\label{eq:square-connection}
\end{equation}
Both probabilities refer to independently retaining each bond with probability $s$.
As $r$ grows, the right-hand side decreases to $\theta(s)$, the probability that the origin lies in an infinite revealed cluster.
On the square lattice, $\theta(s)=0$ for $s\le1/2$, including the critical value~\cite{Harris1960,BollobasRiordan2006}.
Combining Eqs.~\eqref{eq:ppt-proof} and \eqref{eq:square-connection} therefore gives
\begin{equation}
\OP(r)^2\le\Pr_{s,\mathbb Z^2}(0\leftrightarrow\partial\Lambda_r)
\xrightarrow{r\to\infty}0,\qquad s\le\tfrac12.
\label{eq:ppt-long-range}
\end{equation}
The bound is independent of $L$, so it holds after the system-size limit at each fixed $r$.
Since $s=p(2-p)\le1/2$ exactly when $p\le p_0$, this proves the restricted conclusion of Theorem~\ref{thm:main}.
For $p<p_0$, the connection probability and hence the restricted advantage decay exponentially with distance~\cite{AizenmanBarsky1987}.

\subsection{Further consequences for readout}
\emph{The role of Pauli $Y$ terms in a joint observable.}
The score comparison also applies to observables outside the fully PPT class.
\begin{corollary}\label{cor:KH}
For $0<p<1$ and any Hermitian $H$ on the probe space, let $K_H=2^{-M}\sum_{S\subseteq E}H^{T_S}$.
This average removes all Pauli products containing a local $Y$ factor.
Then
\begin{equation}
\Tr BH=\Tr BK_H,\qquad
|\Tr BH|^2\le\Pr_s(i\leftrightarrow j)\,\Tr\rho_AK_H^2.
\label{eq:KH}
\end{equation}
\end{corollary}
Score preservation and cancellation of $Y$ terms hold for every Hermitian $H$.
The moment identities in Eq.~\eqref{eq:conditional-revealed-moments} also apply to every operator containing only $I,X,Z$ factors.
Repeating the first three steps of Eq.~\eqref{eq:ppt-proof} gives Eq.~\eqref{eq:KH}.

A binary readout has an observable $H$ with eigenvalues in $[-1,1]$.
If its score exceeds $\sqrt{\Pr_s(i\leftrightarrow j)}$, Eq.~\eqref{eq:KH} requires $\Tr\rho_AK_H^2>1$.
Some eigenvalues of $K_H$ must then lie outside $[-1,1]$.
The $Y$ terms contribute zero to the score, yet they keep the full observable $H$ within its allowed spectral range.
Such terms can include real matrix products such as $Y_eY_f$.
This describes a necessary feature of observables that beat the restricted bound.

\emph{Relative distinguishability of the same conditional records.}
Fix $p\in[p_*,p_0]$ and write $B_L$ for $B$ in $\Lambda_L$.
Measuring $X$ along a path of length $r$ gives advantage $s^{r/2}>0$.
Thus $\|B_L\|_{\Pall}$ and its system-size limit are positive at every fixed $r$.
Theorem~\ref{thm:main} gives a nonzero long-range joint advantage while the restricted advantage vanishes, so
\begin{equation}
\lim_{r\to\infty}\lim_{L\to\infty}\frac{\|B_L\|_1}{\|B_L\|_{\Pall}}
=\lim_{r\to\infty}\frac{\OG(r)}{\OP(r)}=+\infty.
\label{eq:ratio}
\end{equation}
This ratio quantifies the growing difference between joint and restricted access to the same pair of records.
The system-size limit is taken at fixed $r$ before the distance limit.

\emph{A local readout for stronger records.}
For sufficiently strong records, a simple local measurement already keeps a long-range advantage.
Each probe is measured with three outcomes, identifying $|a_+\rangle$, identifying $|a_-\rangle$, or giving an inconclusive result.
The standard optimal unambiguous discrimination measurement identifies either state without error with probability $1-c=p$~\cite[Sec.~III]{Chefles2000}.
The inconclusive probability is the same for both states.
Independent measurements on the probes therefore reveal physical labels independently with probability $p$.
If the revealed labels contain an $i$--$j$ path, their product gives $T$ exactly.
Otherwise the receiver guesses at random.
All outcomes contribute to the success probability.
This protocol has advantage $\Pr_{p,\Lambda_L}(0\leftrightarrow r\hat x)$, so
\begin{equation}
\begin{aligned}
\mathcal O_{{\rm LOCC},L}(0,r\hat x)&\ge\Pr_{p,\Lambda_L}(0\leftrightarrow r\hat x),\\
\liminf_{r\to\infty}\mathcal O_{\rm LOCC}(r)&\ge\theta(p)^2>0,\qquad p>\tfrac12.
\end{aligned}
\label{eq:locc-strong-record}
\end{equation}
The long-range bound follows from percolation above $1/2$, uniqueness of the infinite cluster and the Harris--FKG inequality~\cite{Kesten1980,AKN1987,BurtonKeane1989,Harris1960,Fortuin1971}.
Here $\theta(p)$ refers to the revealed clusters of this local measurement.
The long-range behavior of the restricted readouts for $p_0<p\le1/2$ remains open.

\subsection{Trees: local measurements attain the joint optimum}
A tree has a unique path $\gamma$ from $i$ to $j$.
Let $d$ be its number of edges.
Multiplying the bond labels along this path gives $T$.
Measuring $X$ on each path probe gives a noisy estimate of its label, with mean $\sqrt s\,b_e$.
Multiplying these outcomes therefore predicts $T$ with advantage $s^{d/2}$.
Joint access to the probes gives the same optimum.
\begin{proposition}\label{prop:tree}
On a finite tree,
\begin{equation}
\OG(i,j)=\OP(i,j)=\mathcal O_{\rm SEP}(i,j)=\mathcal O_{\rm LOCC}(i,j)
=s^{d/2}=\sqrt{\Pr_s(i\leftrightarrow j)}.
\label{eq:tree-readouts}
\end{equation}
\end{proposition}
To see why the joint readout gains nothing further, fix the spin at one site.
The bond labels then determine all other spins.
The uniform spin prior makes these labels independent and equally likely to be $+1$ or $-1$.
Averaging $B=\E_z[T\rho_z]$ thus gives
\begin{equation}
B=\bigotimes_{e\in\gamma}\frac{\sqrt s}{2}X_e
\otimes\bigotimes_{e\notin\gamma}\frac12(I+cZ_e).
\label{eq:tree-factorization}
\end{equation}
Each path factor has trace norm $\sqrt s$.
Each factor off the path is a normalized state and has trace norm 1.
Multiplying these norms gives $\OG=s^{d/2}$, which the local path measurement already attains.
The same result saturates Theorem~\ref{thm:ppt}, since the revealed connection probability on a tree is $s^d$.
For $0<p<1$, the path carries information at every finite distance, and its advantage decays as the path grows.
A readout separation therefore requires a loop.

\subsection{The triangle: an exact finite-graph separation}
The triangle provides two paths between $i$ and $j$.
Label the direct-edge probe $0=ij$ and the two detour probes $1=ik$ and $2=kj$.
Here probe subscripts on $X$ and $Z$ refer to these three qubits.
The labels $b_1$ and $b_2$ are independent and uniform, while $b_0=b_1b_2=T$.
The direct probe records the target itself.
The detour probes give a second estimate of the same target through their product.

A revealed connection uses either the direct edge or both detour edges.
Its probability is $s+(1-s)s^2=s+s^2-s^3$.
For $0<p<1$, the exact readout optima are
\begin{proposition}\label{prop:tri}
\begin{equation}
\begin{aligned}
\mathcal O_{\rm LOCC}=\mathcal O_{\rm SEP}=\OP&=\sqrt{s+s^2-s^3},\\
\OG&=\tfrac12\big[s+\sqrt{s(4-3s)}\big]>\OP.
\end{aligned}
\label{eq:triangle-readouts}
\end{equation}
The restricted optimum is attained by one round of classical communication.
\end{proposition}

\emph{Using the detour to choose the direct-probe measurement.}
First measure $X$ on probes 1 and 2, obtaining $\xi_1,\xi_2=\pm1$.
Their outcomes satisfy $P(\xi_1|b_1)=\tfrac12(1+\xi_1 b_1\sqrt s)$ and the corresponding relation for probe 2.
The two outcomes are independent and uniform.
Bayes' rule then gives
\begin{equation}
P(\xi_1,\xi_2)=\tfrac14,\qquad P(T=\pm1|\xi_1,\xi_2)=\tfrac12(1\pm \xi_1\xi_2\,s).
\label{eq:triangle-posterior}
\end{equation}
The two laboratories send their outcomes to the laboratory holding probe 0.
Only the product $\xi_1\xi_2$ enters its choice of measurement.
Probe 0 is still in $\omega_T$, now with the biased prior in Eq.~\eqref{eq:triangle-posterior}.
The Holevo--Helstrom measurement for these two weighted states~\cite[Theorem~3.4]{Watrous2018} measures
\begin{equation}
H_{\xi_1\xi_2}=\frac{\sqrt s\,X_0+\xi_1\xi_2\,cs\,Z_0}{\sqrt{s+c^2s^2}}.
\label{eq:triangle-adaptive-axis}
\end{equation}
Its two outcomes $\xi_0=\pm1$ are the guesses for $T$.
The $Z_0$ component changes sign with $\xi_1\xi_2$, so the detour controls the direction of the final local measurement.
For either value of $\xi_1\xi_2$, the conditional advantage is $\sqrt{s+c^2s^2}$.
Averaging over all outcomes gives the same advantage, equal to $\sqrt{s+s^2-s^3}$ because $c^2=1-s$.
This reaches the upper bound of Theorem~\ref{thm:ppt} and fixes all three restricted optima in Eq.~\eqref{eq:triangle-readouts}.

Define the single-probe $X$ projectors $\Pi_{\xi_k}=(I+\xi_k X)/2$ for $k=1,2$.
The sequential protocol has effects
\begin{equation}
E_{\xi_1,\xi_2,\xi_0}=\tfrac12(I+\xi_0 H_{\xi_1\xi_2})_0\otimes\Pi_{\xi_1}\otimes\Pi_{\xi_2},
\qquad \xi_1,\xi_2,\xi_0\in\{+1,-1\}.
\label{eq:triangle-povm}
\end{equation}
The last two factors act on probes 1 and 2.
Each step is a complete projective measurement, so these effects are positive and sum to the identity.
The resulting decision observable is
\begin{equation}
H_*=\sum_{\xi_1,\xi_2,\xi_0}\xi_0 E_{\xi_1,\xi_2,\xi_0}
=\frac{\sqrt s\,X_0+cs\,Z_0X_1X_2}{\sqrt{s+c^2s^2}}.
\label{eq:Hstar}
\end{equation}
It has score $\Tr BH_*=\sqrt{s+c^2s^2}$.

\emph{The further gain from a joint readout.}
We obtain the joint optimum directly from $\OG=\|B\|_1$.
Write $A_e=I_e+cZ_e$ on probe $e$.
Averaging the four equally likely choices of $(b_1,b_2)$ gives
\begin{equation}
B=\tfrac18\big(\sqrt s\,X_0A_1A_2+s\,A_0X_1X_2\big).
\label{eq:Btri}
\end{equation}
This operator preserves the $Z$-basis parity of probes 1 and 2.
The subspaces with $Z_1Z_2=+1$ and $Z_1Z_2=-1$ each give a four-dimensional block.
Diagonalizing these blocks gives the nonzero eigenvalues
\begin{equation}
\begin{aligned}
Z_1Z_2=+1&\quad:\quad \lambda=\pm\tfrac14\sqrt{s(4-3s)},\\
Z_1Z_2=-1&\quad:\quad \lambda=\pm\tfrac14s.
\end{aligned}
\label{eq:triangle-spectrum}
\end{equation}
The other four eigenvalues are zero.
Summing the absolute values gives the joint expression in Eq.~\eqref{eq:triangle-readouts}.
Its strict separation from the restricted value follows from
$s(4-3s)-s^2(3-2s)^2=4s(1-s)^3>0$ for $0<s<1$.

\emph{Why the measurement order matters.}
Suppose probe 0 is also measured along $X$, with outcome $\xi_0$.
For this fixed choice of axes, the posterior score is
\begin{equation}
P(\xi_0,\xi_1,\xi_2)\E[T|\xi_0,\xi_1,\xi_2]=\tfrac18(\xi_0\sqrt s+\xi_1\xi_2\,s).
\label{eq:triangle-all-x}
\end{equation}
Since $s<\sqrt s$, its sign is always $\xi_0$.
The detour outcomes change the confidence in $T$ while leaving the best guess equal to the direct-probe outcome.
The resulting advantage is $\sqrt s$.
Using $\xi_1\xi_2$ before measuring probe 0 instead rotates its measurement axis and raises the advantage to $\sqrt{s+c^2s^2}$.
The triangle thus exhibits an adaptive gain within LOCC and a further gain from a joint measurement.

\section{S2. Prediction confidence and information in all outcomes}\label{sec:sm-S2}
After outcome $Y=y$, the posterior mean $m_y=\E[T|Y=y]$ is the difference between the probabilities of the two target values.
It is zero for an outcome uninformative about $T$ and has magnitude 1 when the outcome determines $T$.
The average squared posterior bias is
\begin{equation}
V_Y=\E[m_Y^2],\qquad \E[\operatorname{Var}(T|Y)]=1-V_Y.
\label{eq:posterior-confidence}
\end{equation}
Variance means the mean squared deviation from the mean.
Thus $V_Y$ measures how much of the initial unit variance of $T$ can be predicted from the readout.
The mutual information $I(T{:}Y)$ measures the average reduction in the entropy of $T$, in bits.
Both quantities obey the same connection bound.
\begin{proposition}\label{prop:l2}
For every complete fully PPT measurement on a finite graph,
\begin{equation}
V_Y\le\Pr_s(i\leftrightarrow j),\qquad I(T{:}Y)\le\Pr_s(i\leftrightarrow j).
\label{eq:posterior-information-bounds}
\end{equation}
\end{proposition}

To prove the first bound, apply the transpose average to each effect,
$Q_y=2^{-M}\sum_{S\subseteq E}F_y^{T_S}$.
The $Q_y$ are positive and sum to the identity.
The real product states of the record give $\Tr\rho_zQ_y=\Tr\rho_zF_y$, so the joint statistics of $T$ and $Y$ are unchanged.
The averaging also removes every local Pauli $Y$ term, as in Sec. S1.
Weight these effects by the posterior estimate and by its square,
\begin{equation}
K=\sum_y m_yQ_y,\qquad J=\sum_y m_y^2Q_y.
\label{eq:posterior-observables}
\end{equation}
The score of $K$ and the expectation value of $J$ both reproduce $V_Y$.
Completeness of the measurement also gives the operator inequality
\begin{equation}
\Tr BK=\Tr\rho_AJ=V_Y,\qquad
J-K^2=\sum_y(m_yI-K)Q_y(m_yI-K)\succeq0.
\label{eq:posterior-square-bound}
\end{equation}
Applying Eq.~\eqref{eq:KH} to $K$, which has no local Pauli $Y$ terms, now yields
\begin{equation}
V_Y^2\le\Pr_s(i\leftrightarrow j)\Tr\rho_AK^2
\le\Pr_s(i\leftrightarrow j)\Tr\rho_AJ
=\Pr_s(i\leftrightarrow j)V_Y.
\label{eq:posterior-connection-proof}
\end{equation}
It follows that $V_Y\le\Pr_s(i\leftrightarrow j)$.
The argument at $p=1$ follows directly from the connected-component reasoning in Sec. S1.

To obtain the information bound, define the binary entropy
$h_2(x)=-x\log_2x-(1-x)\log_2(1-x)$, with $h_2(0)=h_2(1)=0$.
The elementary binary-entropy inequality gives
\begin{equation}
1-h_2\!\left(\frac{1+m}{2}\right)\le m^2,\qquad -1\le m\le1.
\label{eq:binary-entropy-bound}
\end{equation}
\equationalias{lem:h2}
Since the two target values have equal prior probabilities,
\begin{equation}
I(T{:}Y)=\E\!\left[1-h_2\!\left(\frac{1+m_Y}{2}\right)\right]
\le V_Y\le\Pr_s(i\leftrightarrow j).
\label{eq:restricted-mutual-information}
\end{equation}
These bounds include the probability-weighted contributions of all outcomes, including rare, confident ones.
For each readout, $(\E|m_Y|)^2\le V_Y$ also recovers Theorem~\ref{thm:ppt} after optimizing the guessing advantage.

\section{S3. The joint readout: proof of the GLOBAL half of Theorem~\ref{thm:main}}\label{sec:sm-S3}

\subsection{From joint prediction to a spin-state overlap}
The joint prediction advantage is the trace norm of $B$.
To express it through the spins, consider the pure state before either system is discarded,
\begin{equation}
|\Psi\rangle=2^{-|V|/2}\sum_z|z\rangle|A_z\rangle,\qquad
B=\operatorname{Tr}_S\!\left[(Z_iZ_j\otimes I)|\Psi\rangle\langle\Psi|\right].
\label{eq:spin-probe-purification}
\end{equation}
Tracing out the probes gives the spin state $\rho_S$.
The states $|\Psi\rangle$ and $(Z_iZ_j\otimes I)|\Psi\rangle$ purify $\rho_S$ and $Z_iZ_j\rho_SZ_iZ_j$.
Their overlap after a unitary $\mathcal U_A$ acts on the probes is $\Tr\mathcal U_AB$.
Maximizing its absolute value gives $\|B\|_1$.
Uhlmann's theorem identifies the same maximum with the root fidelity~\cite[Sec.~3.2]{Watrous2018}, giving
\begin{equation}
\OG=\|B\|_1=F(\rho_S,Z_iZ_j\rho_SZ_iZ_j).
\label{eq:joint-spin-overlap}
\end{equation}
Here $F(\rho,\sigma)=\|\sqrt\rho\sqrt\sigma\|_1$.

\subsection{Dephasing events as vertex parity patterns}
Tracing out each probe induces the bond-dephasing channel of Eq.~\eqref{eq:dephasing}, with $q=p/2$~\cite{Lessa2025}.
Let $N_e=1$ mark a dephasing event on bond $e$, and $N_e=0$ mark its absence.
These bits are independent, with $\Pr(N_e=1)=q$.
Each event applies $Z_uZ_v$ and reverses the $X$-basis spin labels at its two sites.
Starting from all spins in $|+\rangle$, a site has $X$ eigenvalue $-1$ precisely when an odd number of incident events have occurred.

For an edge-bit configuration $n$, write $\partial n$ for its vertex parity pattern,
\begin{equation}
(\partial n)_v=\sum_{e\ni v}n_e\pmod2.
\label{eq:vertex-parity-map}
\end{equation}
An entry 1 therefore marks a site touched by an odd number of selected edges.
Addition of edge or vertex bit configurations, and sums defining vertex parity, are modulo 2 throughout Sec. S3.
Let $t=\partial N$, and let $|t\rangle_X$ be the product state whose spin at $v$ has $X$ eigenvalue $(-1)^{t_v}$.
Then
\begin{equation}
\rho_S=\sum_tP_q(t)|t\rangle_X\langle t|,\qquad
P_q(t)=\sum_{n:\,\partial n=t}q^{|n|}(1-q)^{M-|n|}.
\label{eq:dephasing-parity-distribution}
\end{equation}
Here $|n|$ counts the selected bonds and $M=|E|$.
Different event patterns with the same vertex parities contribute to the same spin state.
The distribution $P_q$ adds their probabilities.

Define the vertex-bit pattern $d$ to be 1 at $i$ and $j$ and 0 elsewhere.
Applying $Z_iZ_j$ changes $t$ to $t+d$.
The two spin states in Eq.~\eqref{eq:joint-spin-overlap} are diagonal in the same basis, so their root fidelity is the overlap of their probabilities,
\begin{equation}
\OG=\sum_t\sqrt{P_q(t)P_q(t+d)}.
\label{eq:spin-parity-affinity}
\end{equation}
This reduces the joint advantage to a classical probability overlap.

\subsection{An auxiliary bit whose output overlap gives the advantage}
On the same finite connected graph, choose an edge pattern $C$ uniformly from those with even parity at every vertex.
Such patterns are unions of closed loops, including the empty pattern.
Call their number $\nu$.
Choose a fixed path $\gamma$ from $i$ to $j$ and represent it by bits equal to 1 on the path and 0 elsewhere.
Choose a uniform bit $U\in\{0,1\}$ independently of $C$ and set
\begin{equation}
x=C+U\gamma,\qquad \partial x=Ud.
\label{eq:auxiliary-input}
\end{equation}
Thus $U=0$ selects even parity everywhere, while $U=1$ selects odd parity at $i$ and $j$.
Observe $y=x+N$, where $N$ is independent of $C$ and $U$ and flips each edge bit with probability $q$.
This is the auxiliary task of inferring $U$ from the noisy edge bits $y$.

Let $Q_u(y)=P(y|U=u)$.
For fixed $u$ and $y$, the compatible noise patterns have vertex parity $\partial y+ud$.
Each compatible input $x$ has weight $1/\nu$, so
\begin{equation}
Q_u(y)=\frac1\nu P_q(\partial y+ud).
\label{eq:auxiliary-output-law}
\end{equation}
Every allowed vertex parity pattern has exactly $\nu$ edge configurations.
Indeed, adding any fixed representative maps the closed-loop patterns onto that class.
Summing over $y$ therefore gives
\begin{equation}
\sum_y\sqrt{Q_0(y)Q_1(y)}
=\sum_t\sqrt{P_q(t)P_q(t+d)}=\OG.
\label{eq:auxiliary-affinity}
\end{equation}
The normalization factor cancels because each vertex pattern has the same number of edge configurations.

For outputs of positive probability, the posterior bias for $U$ is
\begin{equation}
m(y)=P(U=0|y)-P(U=1|y)=\frac{Q_0(y)-Q_1(y)}{Q_0(y)+Q_1(y)},\qquad
P(y)=\frac{Q_0(y)+Q_1(y)}2.
\label{eq:auxiliary-posterior}
\end{equation}
Since $P(y)\sqrt{1-m(y)^2}=\sqrt{Q_0(y)Q_1(y)}$, Eq.~\eqref{eq:auxiliary-affinity} becomes
\begin{equation}
\OG=\E_y\sqrt{1-m(y)^2}\ge1-\E_y[m(y)^2].
\label{eq:auxiliary-overlap}
\end{equation}
\equationalias{prop:ident}
If $y$ leaves $U$ equally likely to be 0 or 1, its overlap factor is 1.
If $y$ determines $U$, the factor is 0.
This exact equality connects uncertainty about the auxiliary bit to the physical joint prediction advantage.

The physical bond labels and the auxiliary bits obey different constraints.
For $b_e=z_uz_v$, the product around any closed loop is $+1$.
For $x_e$, the sum on a star away from $i$ and $j$ is even.
The four-bit comparison below uses this second constraint.

\subsection{Comparing noisy and exact-or-missing observations}
For an input signal $f(x)$, the conditional mean after observing a channel output is its best mean-square estimate.
The variance of that estimate measures how much of the signal's variation is predictable from the observation.
We compare channels $W$ and $E$ with the same input patterns, allowing their outputs to take different forms.
Write
\begin{equation}
W\Ltwo E\quad\text{if}\quad
\operatorname{Var}\E[f(x)|Y_W]\le\operatorname{Var}\E[f(x)|Y_E]
\quad\text{for every input distribution and real }f.
\label{eq:channel-comparison}
\end{equation}
Here $Y_W$ and $Y_E$ are the respective outputs.
Makur and Polyanskiy characterize the standard less-noisy order through $\chi^2$-divergence~\cite[Theorem~1]{MakurPolyanskiy2018}.
Here $\chi^2(P\|Q)=\sum_y[P(y)-Q(y)]^2/Q(y)$ on the support of $Q$.
For an input law $\pi$, choose a real signal $g$ with $\E_\pi g=0$ and a sufficiently small $\epsilon$ so that $\pi_\epsilon(x)=\pi(x)[1+\epsilon g(x)]$ is a probability law.
Then
\begin{equation}
\chi^2(\pi_\epsilon W\|\pi W)
=\epsilon^2\operatorname{Var}_\pi\E[g(x)|Y_W].
\label{eq:chi-square-variance}
\end{equation}
The notation $\pi W$ means the output distribution when the input has law $\pi$.
Applied to all input laws and signals, this identity gives the second-moment comparison in Eq.~\eqref{eq:channel-comparison}.
Input patterns of zero probability can be omitted.
The notation $W(x)$ used below denotes the random output of $W$ for input $x$.

For one bit, $\mathrm{BSC}_c$ denotes a binary symmetric channel that flips it with probability $q=(1-c)/2$.
Let $E_a$ instead reveal the bit exactly with probability $a$ and otherwise report a missing-bit mark.
The standard single-bit comparison gives
\begin{equation}
\mathrm{BSC}_c\Ltwo E_a\quad\Longleftrightarrow\quad a\ge c^2.
\label{eq:single-edge-comparison}
\end{equation}
\equationalias{lem:edge}
This follows directly from the known binary symmetric/erasure channel boundary~\cite[Appendix, Claim~4]{Nair2009}, using the second-moment characterization above.
The literature uses erasure probability $1-a$.
Its condition $1-a\le4q(1-q)$ becomes $a\ge(1-2q)^2=c^2$ in our notation.
The smallest reveal probability in this comparison is therefore $c^2$.
The remaining bits are hidden with probability $1-c^2=s$.

\subsection{Why the comparison can be applied inside a correlated lattice}
The auxiliary bits on different stars are correlated.
Observing other stars changes the probabilities of the allowed input patterns on the star being replaced.
This is why the comparison must hold for every input distribution.

Consider a block of input bits $x_B$.
Let $Y_B$ be its channel output and $Y_{\rm rest}$ all observations outside the block.
The channel noise on the block is independent of all inputs and of the other noise.
For a real target $g$ determined by the input configuration, such as $(-1)^U$, this independence gives
\begin{equation}
\E[g|Y_B,Y_{\rm rest}]
=\E\!\left[\E[g|x_B,Y_{\rm rest}]\,\middle|\,Y_B,Y_{\rm rest}\right].
\label{eq:block-conditional-mean}
\end{equation}
Fixing $Y_{\rm rest}$ turns the inner conditional mean into a function of the block input.
Apply the channel comparison to that function and the resulting conditional input distribution.
If only patterns in a set $\mathscr C_B$ are allowed, the comparison need hold only on that set.
The mean $\E[g|Y_{\rm rest}]$ is shared by both channels.
The ordering of predicted variances therefore also orders the squared posterior means.
Averaging over $Y_{\rm rest}$ yields
\begin{equation}
\E\!\left[\E[g|Y_B,Y_{\rm rest}]^2\right]
\le\E\!\left[\E[g|\widetilde Y_B,Y_{\rm rest}]^2\right],
\label{eq:conditional-replacement}
\end{equation}
\equationalias{lem:replace}
where $\widetilde Y_B$ is the output of the more informative replacement channel, with independent noise.
For $g=(-1)^U$, replacing blocks in this way can only increase $\E m^2$.
Equation~\eqref{eq:auxiliary-overlap} then gives a lower bound on the physical joint advantage.

\subsection{The four-bit comparison}
The gain comes from the even-parity rule on a four-edge star [Fig.~\ref{fig:mechanism}(b)].
Once two bits are known, the other two must be equal or opposite.
A known flip turns the latter case into two copies of the same unknown bit.
We use this redundancy to compare the whole star with an observation that reveals each bit less often than the single-edge bound requires.

The allowed star inputs are the eight patterns in
\begin{equation}
\mathscr C=\left\{x\in\{0,1\}^4:\sum_{k=1}^4x_k=0\pmod2\right\}.
\label{eq:star-inputs}
\end{equation}
Write $W_c$ for four independent bit-flip channels, each with flip probability $(1-c)/2$.
Write $E_\tau$ for four independent exact-or-missing observations, each revealing its bit with probability $\tau$.
The input bits can be correlated.
For any input distribution $\pi$ on $\mathscr C$ and any real signal $f(x)$, define
\begin{equation}
\mathcal V={\rm Var}_\pi f,\qquad I_{\mathcal W}={\rm Var}\,\E[f|\mathcal W(x)].
\label{eq:star-predicted-variance}
\end{equation}
Here $\mathcal V$ is the signal variance before observation and $I_{\mathcal W}$ is the part predicted by channel $\mathcal W$.
The symbol $I_{\mathcal W}$ denotes a variance, rather than mutual information.

At $p=p_*$, we can lower the reveal probability to
\begin{equation}
\tau_*=\tfrac12-10^{-7},\qquad 1-\tau_*>\tfrac12.
\label{eq:hidden-density}
\end{equation}
\begin{lemma}\label{lem:block}
Let $p_*=0.29181$ and $c_*=1-p_*$. For every input distribution on $\mathscr C$ and every real function $f$,
\begin{equation}
{\rm Var}\,\E[f\,|\,W_{c_*}(x)]\ \le\ {\rm Var}\,\E[f\,|\,E_{\tau_*}(x)]-4\times10^{-6}\,{\rm Var}f .
\label{eq:block}
\end{equation}
The input distribution may assign zero probability to some patterns.
\end{lemma}
We prove the comparison by keeping track of the uncertainty left after some bits are known.
For a subset $R\subseteq\{1,2,3,4\}$, let $x_R$ be the bits at those positions and put
\begin{equation}
H_R=\E\,{\rm Var}(f|x_R),\qquad
H_1=\sum_{|R|=1}H_R,\qquad H_2=\sum_{|R|=2}H_R.
\label{eq:star-residual-variance}
\end{equation}
Thus $H_R$ is the average remaining signal variance after those bits are revealed.
The sums $H_1$ and $H_2$ contain four and six terms, respectively.
Three known bits fix the fourth by parity, so $H_R=0$ for $|R|\ge3$.
Conditioning on which bits $E_\tau$ reveals gives
\begin{equation}
\begin{aligned}
I_{E_\tau}&=\ell(\tau)\mathcal V-w_1(\tau)H_1-w_2(\tau)H_2,\\
H_1&\le2H_2,
\end{aligned}
\label{eq:star-erasure-variance}
\end{equation}
\equationalias{lem:code}
where $\ell(\tau)=1-(1-\tau)^4$, $w_1(\tau)=\tau(1-\tau)^3$, and $w_2(\tau)=\tau^2(1-\tau)^2$.
These are the probability of revealing at least one bit and the probabilities of revealing a specified single bit or pair.

The inequality in Eq.~\eqref{eq:star-erasure-variance} follows by comparing pairs of allowed input patterns.
Write $\pi_x$ for the probability of pattern $x$.
Within a group $\mathcal B$ of patterns sharing given revealed values, its contribution to the residual variance is
\begin{equation}
\pi(\mathcal B){\rm Var}(f|x\in\mathcal B)
=\sum_{\{x,x'\}\subseteq\mathcal B}
\frac{\pi_x\pi_{x'}}{\pi(\mathcal B)}[f(x)-f(x')]^2.
\label{eq:star-variance-pairs}
\end{equation}
The sum counts each distinct pair once, and groups of zero probability contribute zero.
Two even-parity patterns differ in either two or four positions.
If they differ in two positions, they agree in the other two.
They then contribute to two terms in $H_1$ and one term in $H_2$.
The group for the latter contains exactly these two patterns.
Each group for $H_1$ has at least their combined probability, so either contribution is no larger than the one in $H_2$.
Patterns differing in all four positions contribute to neither sum.
Summing proves $H_1\le2H_2$ for every $\pi$ and $f$.

Now set $a=c^2$, the reveal probability required by the single-edge comparison.
Consider two noisy copies of one bit.
They agree with probability $(1+a)/2$, independently of the input bit.
Upon agreement, their common value has correlation $2c/(1+a)$ with that bit.
Disagreement gives no information about it.
Equation~\eqref{eq:single-edge-comparison} therefore bounds the predictable fraction of any signal variance by $2a/(1+a)$.
Two independent exact-or-missing copies instead determine the bit with probability $2a-a^2$.
Their gain in predictable variance is at least
\begin{equation}
\Delta_2\,{\rm Var}f,\qquad
\Delta_2=(2a-a^2)-\frac{2a}{1+a}
=\frac{a^2(1-a)}{1+a}.
\label{eq:repeated-pair-gain}
\end{equation}
Here $f$ is any real function of the repeated bit.

To use this gain on the star, choose a pair $R$ and first replace its two noisy observations by reveals at rate $a$.
Every replacement can only increase the predictable variance by Eq.~\eqref{eq:conditional-replacement}.
With probability $a^2$, both bits in $R$ are revealed.
The remaining pair is then a repeated bit, up to a known flip.
Replacing that pair gains at least $\Delta_2 H_R$ on average within this event.
Thus $I_{E_a}-I_{W_c}\ge a^2\Delta_2 H_R$ for each of the six choices of $R$.
Averaging these six bounds gives
\begin{equation}
I_{W_c}\le I_{E_a}-\kappa H_2,\qquad
\kappa=\frac{a^2\Delta_2}{6}=\frac{a^4(1-a)}{6(1+a)}.
\label{eq:star-repeated-gain}
\end{equation}
\equationalias{lem:rep}

For input distributions with small $H_2$, we also need a bound independent of this repeated-bit gain.
Raginsky's variance characterization identifies the largest predictable fraction with the $\chi^2$ contraction coefficient~\cite{Raginsky2016}.
This coefficient is the largest fraction of $\chi^2$-divergence retained by the channel.
The Dobrushin bound limits that coefficient by the largest total-variation distance between two channel output distributions.
For two distributions $P,Q$, this distance is $\tfrac12\sum_y|P(y)-Q(y)|$.
Two allowed star inputs differing in two positions have output distance $c$.
Those differing in four positions have distance $c(3-c^2)/2$.
Consequently, for every input distribution,
\begin{equation}
I_{W_c}\le D\mathcal V,\qquad D=\frac{c(3-c^2)}{2}.
\label{eq:star-contraction}
\end{equation}
\equationalias{lem:dob}

We combine Eqs.~\eqref{eq:star-repeated-gain} and \eqref{eq:star-contraction} with weights $1-\beta$ and $\beta$, where $0\le\beta\le1$.
Subtracting the resulting upper bound from $I_{E_\tau}$ in Eq.~\eqref{eq:star-erasure-variance} gives
\begin{equation}
I_{E_\tau}-I_{W_c}\ge \mathsf A_\beta \mathcal V+\mathsf B_\beta H_1+\mathsf C_\beta H_2,
\label{eq:star-combined-bound}
\end{equation}
with coefficients
\begin{equation}
\begin{aligned}
\mathsf A_\beta&=\ell(\tau)-(1-\beta)\ell(a)-\beta D,\\
\mathsf B_\beta&=(1-\beta)w_1(a)-w_1(\tau),\\
\mathsf C_\beta&=(1-\beta)[w_2(a)+\kappa]-w_2(\tau).
\end{aligned}
\label{eq:star-coefficients}
\end{equation}
If $\mathsf B_\beta\le0$, then $H_1\le2H_2$ bounds the right-hand side from below by $\mathsf A_\beta \mathcal V+(2\mathsf B_\beta+\mathsf C_\beta)H_2$.
It follows that
\begin{equation}
\left.
\begin{aligned}
\mathsf A_\beta&\ge\delta>0,\\
\mathsf B_\beta&\le0,\\
2\mathsf B_\beta+\mathsf C_\beta&\ge0
\end{aligned}\right\}
\quad\Longrightarrow\quad I_{E_\tau}-I_{W_c}\ge\delta \mathcal V.
\label{eq:star-comparison-criterion}
\end{equation}
\equationalias{prop:explicit}

At $p=p_*$, set $c=c_*$ and $\tau=\tau_*$ and choose the mixing weight $\beta=0.01434$.
Substitution into Eq.~\eqref{eq:star-coefficients} in exact rational arithmetic gives
\begin{equation}
\mathsf A_\beta>5\times10^{-6},\qquad
\mathsf B_\beta<-10^{-3},\qquad
2\mathsf B_\beta+\mathsf C_\beta>4\times10^{-6}.
\label{eq:star-coefficient-check}
\end{equation}
\equationalias{cor:point}
The finite decimals and $\tau_*=\tfrac12-10^{-7}$ specify exact rational values.
Taking $\delta=4\times10^{-6}$ in Eq.~\eqref{eq:star-comparison-criterion} proves Lemma~\ref{lem:block}.
The star constraint has thus lowered the required reveal probability below $1/2$.
This is the step that opens the separation window.

\subsection{From star comparisons to hidden paths}
Color the vertices of $\Lambda_L$ black and white like a checkerboard.
Assign every edge to its black vertex, so the resulting stars do not share edges.
A four-edge star centered at a vertex other than $i$ and $j$ has even auxiliary parity.
Its input therefore lies in $\mathscr C$.
Stars at the boundary may have fewer than four edges.
Stars centered at $i$ or $j$, when these vertices are black, need not have even parity.
We treat the edges of these stars individually.

Fix $p=p_*$ and apply the replacements one star at a time.
Observations elsewhere can change the probabilities of the allowed patterns on a star.
Lemma~\ref{lem:block} applies to every such conditional distribution.
The observation noise on each star is independent of the inputs and the noise on other stars.
Equation~\eqref{eq:conditional-replacement}, with $g=(-1)^U$, therefore allows replacement by $E_{\tau_*}$ on every four-edge star away from $i$ and $j$.
Use the single-edge comparison at reveal probability $c_*^2$ on the remaining edges.

Let $m_{\rm er}$ be the posterior difference between the probabilities of $U=0$ and $U=1$ after all replacements.
We have $\E m^2\le\E m_{\rm er}^2$.
Each edge is now independently revealed or hidden.
This independence concerns which bits are missing, even though the bit values remain correlated.
If a hidden path joins $i$ and $j$, flip all its bits [Fig.~\ref{fig:mechanism}(c)].
The flip preserves every revealed value and the even parity at all other vertices.
It changes $U$ to $1-U$.
Because the allowed input patterns have equal prior weight, the two values of $U$ remain equally likely.

If no hidden path joins $i$ and $j$, take the set of vertices reachable from $i$ along hidden edges.
It contains $i$ and excludes $j$.
Sum their vertex parities modulo 2.
Internal edges contribute twice and cancel, leaving only the edges that cross out of this set.
All these crossing edges are revealed.
Their known values therefore determine $U$.
Using Eq.~\eqref{eq:auxiliary-overlap}, we obtain
\begin{equation}
\OG\ge1-\E m_{\rm er}^2=\Pr_{\bm h}(i\leftrightarrow j),\qquad
h_e=\begin{cases}
1-\tau_* & \text{on four-edge stars away from $i,j$},\\
s_* & \text{on the remaining edges},
\end{cases}
\label{eq:hidden-path-bound}
\end{equation}
where $s_*=1-c_*^2=p_*(2-p_*)$.
Here $\Pr_{\bm h}$ is the connection probability with independent hidden-edge probabilities $h_e$.

To avoid the box boundary, use only edges of $\Lambda_{L-1}$ inside $\Lambda_L$.
Every such edge belongs to a four-edge star in the larger box.
Among them, only edges assigned to black stars centered at $i$ or $j$ have the smaller hidden probability $s_*$.
There are at most eight such edges.
The event that they are all hidden has probability at least $s_*^8$.
Conditional on this event, all other edges in the smaller box are independently hidden with probability $1-\tau_*$.
The required edges are certainly present, which can only increase the connection probability relative to using $1-\tau_*$ on every edge.
Keeping just this event in the probability sum yields, for $L\ge r+1$,
\begin{equation}
\mathcal O_{{\rm G},L}(0,r\hat x)\ge s_*^8\,\Pr_{1-\tau_*,\Lambda_{L-1}}(0\leftrightarrow r\hat x),
\label{eq:global-box}
\end{equation}
at $p=p_*$.
The factor $s_*^8$ is independent of distance.

Take $L\to\infty$ at fixed $r$ first.
The connection probability on the right increases to its value on the infinite square lattice.
At hidden probability $h=1-\tau_*>1/2$, a vertex belongs to an infinite hidden cluster with positive probability $\theta(h)$~\cite{Kesten1980}.
The infinite cluster is unique~\cite{AKN1987,BurtonKeane1989}.
Thus, if both $i$ and $j$ belong to an infinite cluster, they are connected.
The Harris--FKG inequality bounds the probability of this joint event from below by $\theta(h)^2$~\cite{Harris1960,Fortuin1971}.
Equation~\eqref{eq:global-box} consequently gives
\begin{equation}
\liminf_{r\to\infty}\OG(r)\ge s_*^8\theta(1-\tau_*)^2>0
\qquad(p=p_*).
\label{eq:joint-long-range-proof}
\end{equation}

\subsection{Monotonicity in \texorpdfstring{$p$}{p}}
The receiver can turn a stronger record into a weaker one by applying a local channel to each probe before measuring it.
Any success probability available from the weaker record is therefore also available from the stronger one.
\begin{proposition}\label{prop:mono}
On every finite graph, for fixed $i,j$ and $0<p_1\le p_2\le1$,
\begin{equation}
\mathcal O_{\mathcal M}(i,j;p_2)\ge\mathcal O_{\mathcal M}(i,j;p_1),\qquad
\mathcal M\in\{{\rm G},{\rm SEP},\Pall,{\rm LOCC}\}.
\label{eq:record-monotonicity}
\end{equation}
\end{proposition}
\begin{proof}
For $p_1=p_2$, use the identity channel.
Otherwise set $c_k=1-p_k$, so $c_1>c_2\ge0$.
Write $|a_\pm(p_k)\rangle$ for the two probe states at strength $p_k$.
Choose normalized auxiliary states $|g_\pm\rangle$ with overlap $c_2/c_1$.
The local map
\begin{equation}
|a_\pm(p_2)\rangle\longmapsto |a_\pm(p_1)\rangle|g_\pm\rangle,
\qquad \langle g_+|g_-\rangle=\frac{c_2}{c_1},
\label{eq:local-record-degradation}
\end{equation}
preserves the norms and overlap of the two input vectors.
It can therefore be implemented by an isometry with a local auxiliary system.
Discarding that system gives a channel $\Lambda_e$ taking either stronger probe state to its weaker counterpart.
This construction also works for $p_2=1$, where the two stronger states are orthogonal.

Apply $\Lambda=\bigotimes_e\Lambda_e$ to the whole record.
A measurement effect $F_y$ on the weaker record becomes $\Lambda^\dagger(F_y)$ on the stronger record.
The adjoint is defined by $\Tr[\Lambda(\rho)F]=\Tr[\rho\Lambda^\dagger(F)]$, so the outcome probabilities agree.
These local channels preserve separable effects and can be prepended to any LOCC protocol.
They also preserve the fully PPT condition because
\begin{equation}
[\Lambda^\dagger(F)]^{T_S}=\widetilde\Lambda^\dagger(F^{T_S}).
\label{eq:degradation-ppt}
\end{equation}
Here $\widetilde\Lambda$ is obtained by complex conjugating the local Kraus matrices of $\Lambda$ on the probes in $S$.
It is again a quantum channel, so its adjoint takes positive operators to positive operators.
Thus every allowed measurement on the weaker record has an allowed counterpart on the stronger one with the same statistics.
\end{proof}
The same monotonicity holds after taking the system-size limit at fixed distance.
Hence Eq.~\eqref{eq:joint-long-range-proof} extends the positive joint lower bound to all $p\in[p_*,1]$.
The restricted advantage vanishes for $0<p\le p_0$ by Sec.~S1.
The overlap $[p_*,p_0]$ proves the separation in Theorem~\ref{thm:main}.

\subsection{Product marginals and path information}
We now return to the physical bond labels $b_e=z_uz_v$.
Their loop constraints correlate the unconditioned probe record.
On bonds without a closed loop, these labels can instead be chosen independently.
\begin{proposition}\label{prop:forest}
Let $\mathcal F\subseteq E$ be a set of bonds containing no closed loop.
The labels $(b_e)_{e\in\mathcal F}$ are independent and equally likely to be $+1$ or $-1$.
The reduced probe record on these bonds is
\begin{equation}
\rho_{A,\mathcal F}=\bigotimes_{e\in\mathcal F}
\begin{pmatrix}1-p/2&0\\0&p/2\end{pmatrix}_e.
\label{eq:forest-record}
\end{equation}
In particular, any two distinct probes have a product reduced state when the graph has no self-bonds or multiple bonds between the same sites.
\end{proposition}
\begin{proof}
The bonds in $\mathcal F$ form a forest, a collection of trees and isolated vertices.
Choose any bond-label assignment and one spin in each connected component.
Propagating along the tree bonds then fixes all other spins without encountering a consistency condition.
Every label assignment arises from the same number of spin configurations and hence has the same probability.
Averaging $|a_b\rangle\langle a_b|$ over $b=\pm1$ cancels its off-diagonal entries.
The independent label averages give Eq.~\eqref{eq:forest-record}.
\end{proof}
This product structure refers to the record before conditioning on $T$.
Along an $i$--$j$ path $\gamma$, the product $\prod_{e\in\gamma}b_e=T$.
Fixing $T$ therefore constrains those labels together, even when the path lies in a tree.
The probes on such a path carry predictive information, as shown by the local protocol in Eq.~\eqref{eq:tree-readouts}.
Loops are needed for the separation between readouts, rather than for path information itself.

\subsection{Statistical-mechanics interpretation and the numerical threshold}
The wider separation region discussed in the main text is suggested by the known random-bond Ising representation of the dephased spin state~\cite[Sec.~III.2]{Lessa2025}.
Here we state that mapping in our notation, with dephasing rate $q=p/2$.
On a square box, place classical Ising variables $\sigma_f=\pm1$ at the centers of its faces.
Fix the variable for the exterior face to $\sigma_{\rm out}=+1$.
A dephasing pattern $N$ sets the coupling signs $\eta_e=(-1)^{N_e}$, so negative signs occur independently with probability $q$.
For the two faces $f,g$ on either side of edge $e$, define
\begin{equation}
\mathcal Z_\eta(\mathcal K)=\sum_{\sigma:\,\sigma_{\rm out}=1}
\exp\!\left(\mathcal K\sum_e\eta_e\sigma_f\sigma_g\right),\qquad e^{-2\mathcal K}=\frac{q}{1-q}.
\label{eq:rbim-partition}
\end{equation}
The dimensionless coupling $\mathcal K$ and sign disorder are thus matched on the Nishimori line~\cite{Nishimori1981}.
A boundary edge has the exterior face on one side.

Event patterns with the same vertex parity differ by closed loops.
On the planar box these loops are generated by changing the face variables.
Their dephasing probabilities therefore give the partition-function weight
\begin{equation}
P_q(\partial N)=[q(1-q)]^{M/2}\mathcal Z_\eta(\mathcal K).
\label{eq:rbim-weight}
\end{equation}
Fixing the exterior variable counts each loop pattern once.
Changing the vertex parity at $i$ and $j$ corresponds to reversing $\eta_e$ along an $i$--$j$ path $\gamma$.
Call the resulting signs $\eta^\gamma$.
The same relation gives $P_q(\partial N+d)=[q(1-q)]^{M/2}\mathcal Z_{\eta^\gamma}(\mathcal K)$.
The ratio of the two partition functions is independent of which path is chosen.
Its dimensionless defect free energy is $\Delta\mathcal F_{ij}=-\ln(\mathcal Z_{\eta^\gamma}/\mathcal Z_\eta)$.
Using Eq.~\eqref{eq:spin-parity-affinity} and the known mapping gives
\begin{equation}
\OG(i,j)=\E_q e^{-\Delta\mathcal F_{ij}/2}
\ge e^{-\E_q\Delta\mathcal F_{ij}/2}.
\label{eq:rbim-affinity}
\end{equation}
The inequality is Jensen's.
The average $\E_q$ runs over the independent coupling signs.
The partition function already sums over the Ising variables at fixed signs.
If the mean defect cost stays bounded as $i$ and $j$ move apart, Eq.~\eqref{eq:rbim-affinity} keeps the joint advantage bounded away from zero.
All identities are taken first at fixed finite box and distance, with the limit order of the main text.

Wan, Dai and Zhu estimated the Nishimori transition through transfer matrices and finite-size scaling~\cite{WanDaiZhu2026}.
Their disorder probability is our $q$, giving $q_N=0.1092212(4)$ and hence $p_G=2q_N=0.2184424(8)$.
This literature estimate suggests separation for $p_G<p\le p_0$.
The analytic proof above establishes the certified interval $[p_*,p_0]$ within that broader region.

\endgroup
\end{document}